\documentclass[prl,showkeys,twocolumn]{revtex4}
\usepackage{graphicx,amssymb,amsmath}
\usepackage{hyperref}
\usepackage{enumitem}
\usepackage{empheq}
\usepackage{color,soul}
\usepackage{mathrsfs}
\usepackage{afterpage}
\usepackage{url}

\usepackage{caption}

\begin{document}

\title{Revisiting the coupling between accessibility and population growth}

\author{Valerio Volpati}
\affiliation{Capital Fund Management, 23-25, Rue de l'Universit\'e 75007 Paris, France}
\affiliation{Chair of Econophysics and Complex Systems, \'Ecole polytechnique, 91128 Palaiseau Cedex, France}

\author{Marc Barthelemy}
\email{marc.barthelemy@ipht.fr}
\affiliation{Institut de Physique Th\'eorique, Universit\'e Paris Saclay, CEA, CNRS, F-91191 Gif-sur-Yvette, France}
\affiliation{Centre d'Analyse et de Math\'ematique Sociales, (CNRS/EHESS) 54, Boulevard Raspail, 75006 Paris, France}

\begin{abstract}

  The coupling between population growth and transport accessibility has been an elusive problem for more than 60 years now. Due to the lack of theoretical foundations, most of the studies that considered how the evolution of transportation networks impacts the population growth are based on regression analysis in order to identify relevant variables. The recent availability of large amounts of data allows us to envision the construction of new approaches for understanding this coupling between transport and population growth. Here, we use a detailed dataset for about 36000 municipalities in France from 1968 until now.  In the case of large urban areas such as Paris, we show that growth rate statistical variations decay in time and display a trend towards homogeneization where local aspects are less relevant. We also show that growth rate differences due to accessibility are very small and can mostly be observed for cities that experienced very large accessibility variations. This suggests that the relevant variable for explaining growth rate variations is not the accessibility but its temporal variation.  We propose a model that integrates the stochastic internal variation of the municipalitie's population and an inter-urban migration term that we show to be proportional to the accessibility variation and has a limited time duration. This model provides a simple theoretical framework that allows to go beyond econometric studies and sheds a new light on the impact of transportation modes on city growth.

\end{abstract}


\maketitle


\section{Introduction}

A central point in urbanism is the relation between accessibility and population growth rate. In other words, the question is to understand and to quantify the effect of a growing connectivity -- or improving accessibility -- on the population growth of cities inside an urban area. A large body of literature discuss the interaction (and coevolution) between transportation infrastructure and suggest that an increase in population (and/or jobs, land use, development etc.) is expected for cities that are affected by this improved connectivity (see for example \cite{Anas:1998,Baum:2007,Xie:2011,Sun:2015,Mimeur:2018,Bottinelli:2019}). In general, the local population density is in general strongly affected by economic factors \cite{Glaeser:2000}, and it is expected that the population growth rate will be larger in high accessibility areas. The largest part of evidences to support this fact comes from econometric approaches, starting with the seminal paper on accessibility \cite{Hansen:1959}. Accessibility can be measured by different variables \cite{Ingram:1971,Jones:1981,Koenig:1980,Handy:1997,Geurs:2004} and quantifies in general how easy it is to go around. Using population density as dependent variables, and accessibility measures amongst the explanatory variables, a regression analysis allows to determine the relative impact, or explanatory power of different transport modes \cite{Koopmans:2012} or of the temporal component of a given mode \cite{Duranton:2012, Garcia-Lopez:2015, Kotavaara:2011, Mayer:2015}. However, these approaches are exclusively empirical and clear theoretical foundations on accessibility and its link with population growth are missing despite their critical importance for identifying the principles that govern urban expansion \cite{Seto:2012} and the evolution of infrastructures \cite{Cats:2020}. In several models of urban dynamics (eg. in Land Use Transport Interaction or LUTI models), accessibility is used as a key factor that determines growth \cite{Batty:2007, Iacono:2008}, but here also there are no clear theoretical justifications.

An important merit of the accessibility measure introduced in the paper \cite{Hansen:1959} is its predictive power with respect to land use development: the development ratio of an area (defined as the ratio of the actual development of the area with respect to its probable development) depends as a power law of the accessibility with an exponent of order $3$ \cite{Hansen:1959} for the urban area of Washington and for various types of accessibilities. These striking results constitute the basis for accessibility studies, and triggered a wealth of studies and works in econometric regression analysis that are based on gravity law measures. In particular, it has been claimed \cite{Song:1996} that gravity-type accessibility measures have the largest explanatory power, performing better then 8 different types of accessibility measures. In several cases, the regression analysis is performed in terms of population density, and not in terms of used land \cite{Kotavaara:2011, Koopmans:2012, Song:1996}. In \cite{Kotavaara:2011}, using several measures of potential accessibility, it is shown how in the Netherlands the development of the railway network had a positive effect on population growth, especially at the beginning of the 20th century when industrialization took off. A similar study is carried out in \cite{Koopmans:2012}, where the effect of the road and the railway networks on growth are studied for Finland in the period 1970-2007. The results of the study point out that in Finland the population distribution is mainly concentrated in areas with high road-based gravity accessibility. The largest statistical significant effect for the railway network is obtained when the time to the nearest station is used in combination with the potential accessibility by road network alone. In particular, the conclusion of the study is that railways had a positive effect on growth mostly at a local level, and mainly in the 1970s and in the period 2000-2007. Other studies with simpler accessibility measures but with more elaborate econometric analysis can also be found. For example, in the study \cite{Duranton:2012}, the authors study the impact of interstate highways on the growth of cities in the U.S between 1983 and 2003 and, using a two-stages regression analysis, concluded that a $1\%$ increase in the stock of highway of a city causes about $0.15\%$ increase of its employment, over the 20 years period considered. Along these lines, Mayer and Trevien \cite{Mayer:2015} focused on the Paris Metropolitan Area and evaluated the impact of the Réseau Express Régional (RER) on growth in the period 1975-1990. Using various regression analysis, they found a strong impact of the RER on the number of jobs (between $7\%$ and $11\%$), but didn't find a significant impact on population density. The impact of the RER in the Paris Metropolitan Area was also studied in \cite{Garcia-Lopez:2015}, where instead a positive impact on population density is found: with each additional kilometer a municipality is located closer to a RER station, employment increases by $2\%$ and population increases by $1\%$. These effects are considerably stronger when a municipality is located less than 13 km from a RER station. In this case, the growth of employment and population is found to be of $12\%$ and $8\%$ per additional kilometer.

Most of these studies are based on a regression analysis of a quantity such as the population with accessibility measures. A first remark is that the impact of accessibility is usually very weak and the effect difficult to exhibit from empirical observations, in sharp constrat with the original observations in \cite{Hansen:1959} and pointing to the need for an explanation of this apparent discrepancy. Another important point is the lack of theoretical foundations and the lack of even a simple toy model that could point to the type of relation between these different variables. The recent availability of new sources of data allow us to envision the construction of such models and to test their predictions against empirical observations \cite{Barthelemy:2019}. Here, we will present an analysis on the evolution of population of the 10 largest urban areas in France from 1968 to 2014 and we show that it is the accessibility variation that impacts the growth rate, an effect that decays relatively quickly in time and in space, as we show for the Paris urban area. Also in order to observe this effect we have to focus on the subset of cities that experienced a very large accessibility variation. This led us to propose a simple model able to explain empirical observations about the population growth.

\section*{Growth rates and accessibility}

\subsection*{The dataset}

We will use data available from the National Institute of Statistics and Economic Studies (INSEE) \cite{Insee}. The dataset contains the population of each municipality in France for the years 1968, 1975, 1982, 1990, 1999, and all years from 2006 to 2014. The number of municipalities is not the same every year, due to merging and separation of administrative units: it fluctuates from a minimum of 35,891 municipalities in 1982, to a maximum of 37,727 in 1968.  Here, we focus on the 35,513 municipalities that are present in the INSEE list for all the years in the database, and we will concentrate on a subset of 4,457 municipality belonging to the 10 largest urban areas in France. We will consider more specifically the urban area of Paris (ie. the region Ile-de-France, see the Fig. S1 for a map). For each municipality $i$, we consider its population $P_i(t)$ for a given year $t$, and its population $P_i(t + \delta t)$ for the following available year. We define the growth rate at time $t$, $g_i(t)$  as
\begin{equation}
\label{eq:gr_definition}
g_i (t) = \frac{P_i(t+\delta t) - P_i(t)}{P_i(t) \delta t}.
\end{equation}
This is the standard definition for growth rates, and we observe that they depend very weakly on the population (see SI Fig S2). It takes two years to define a growth rate and from the 14 years of available population data, we can compute 13 growth rates. 

\subsection*{Homogeneization in large urban areas}

\begin{figure*}[ht!]
  \includegraphics[scale=0.35]{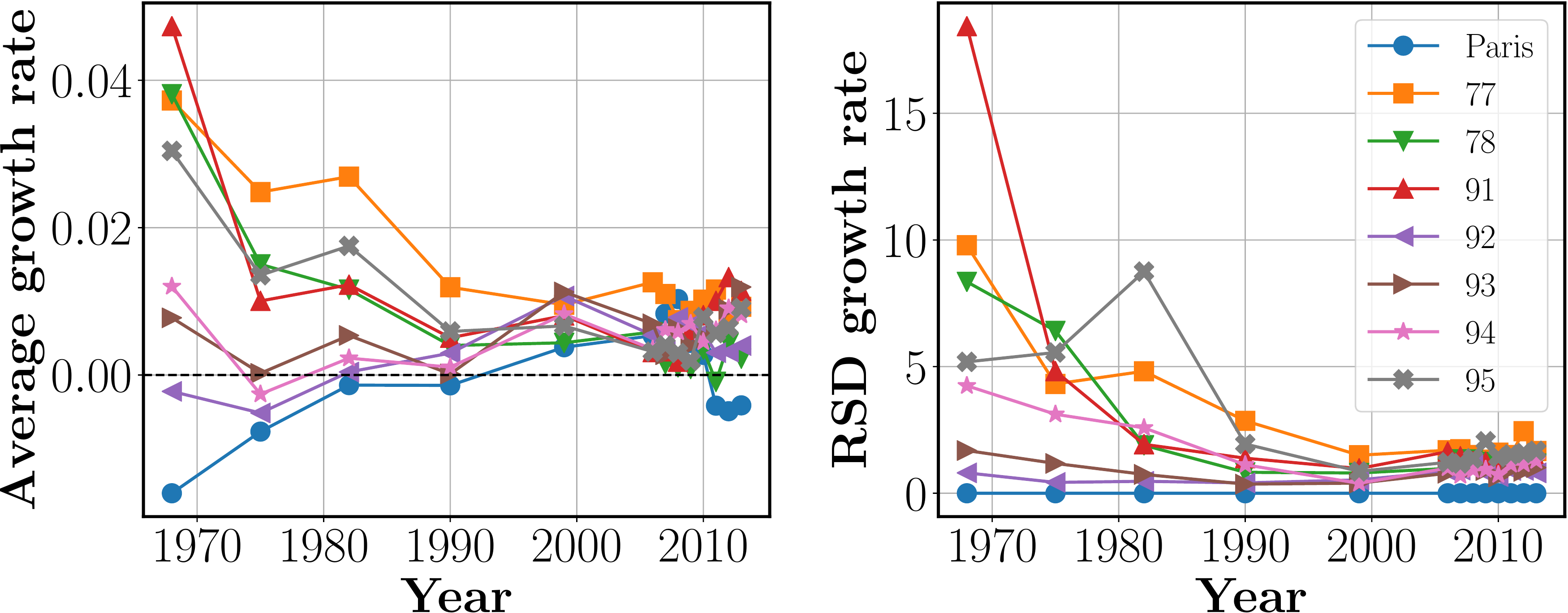}
    \includegraphics[scale=0.35]{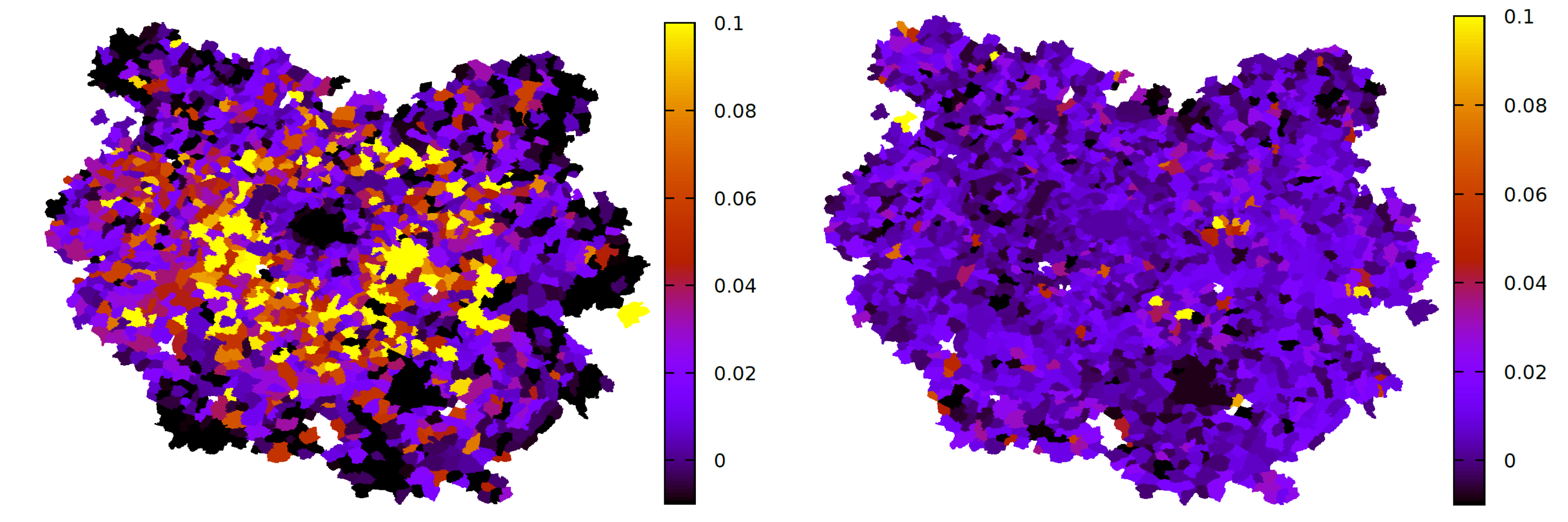}
\caption{(Top) On the left panel, we show the aggregate growth rates in the Ile-de-France departments comprised in the urban area of Paris. For earlier periods, the growth rate of the city of Paris (in blue), is smaller than the growth rates of the other departments of Ile-de-France: Seine-et-Marne (77), Yvelines (78), Essonne(91), Hauts-de-Seine(92), Seine-Saint-Denis (93), Val-de-Marne (94) and Val-d'Oise (95).  This phenomenon can be interpretated as \emph{suburbanization} and \emph{urban sprawl} in the Paris region. At later periods, the aggregate growth rates of different departments become similar pointing to a \emph{homogeneization} phenomenon.  In the right panel we show the relative standard deviation (RSD), defined as the standard deviation of the growth rates of all cities in a region divided by the average. The RSD decreases with time for all the departments, showing that the aggregate growth rate, computed in the previous figure, has always smaller fluctations around its average. (Bottom) Growth rates in Ile-de-France for the period (left) 1968-1975 and (right) for the period 2006-2014. Homogeneization can be seen visually here by looking at these map of growth rates. While in the period 1968-1975, growth rates are more heterogeneous in space, the spatial distribution in the period 2006-2014 looks more homogeneous.}
\label{fig:average}
\end{figure*}

We first focus on the time evolution of growth rate in large urban areas such as the Paris urban area (the region Ile-de-France, see SI for a map and more details) and we will define the aggregate growth rate of a region, or for a generic collection of municipalities, as
\begin{equation}
g_{\alpha}(t) = \frac{P_{\alpha}(t+\delta t)-P_{\alpha}(t)}{P_{\alpha}(t) \delta t}.
\end{equation}
where $P_{\alpha} = \sum_{i \in \alpha} P_i$ is the total population of the region $\alpha$. In the case of the Paris urban area, there are 8 `departments' (Paris proper, Seine-et-Marne, Yvelines, Essonne, Hauts-de-Seine, Seine-Saint-Denis, Val-de-Marne, and Val-d'Oise), over each of which we aggregate cities. In the left panel of Fig. \ref{fig:average} we report the observed aggregate growth rates for all the departments in Ile-de-France in the considered time window.
Despite the fact that the region has shown a marked growth in the time window considered here with an increase of its total population from about $9.5$ millions in 1968 to $12.4$ million in 2014, the city proper of Paris (called `intra-muros' in french) displays the opposite trend: in 1968 Paris hosted about $2.59$ million people and in 2014 about $2.22$ million inhabitants. This is a phenomenon of sub-urbanization where Paris proper has a growth rates lower than those of surrounding departments, and points to urban sprawl in this urban area -- an expansion of human population away from central urban areas, that has become prominent in western urban areas since the 90s (see for example \cite{Champion:2001,Antrop:2004,Fielding:1982,Geyer:1993}). Another significant feature of Fig. \ref{fig:average} is the fact that aggregate growth rates of different departments converge to the same rate value, typically between $0\%$ and $2\%$ yearly. This observation is a first facet of a process of \emph{homogeneization}, according to which peculiarities of different regions (or cities) have become less relevant as years go by. Homogeneization in this Ile-de-France region can also be observed visually in Fig.~\ref{fig:average}(bottom) where we compare maps of growth rates in the period 1968-1975, versus the map of the period 2006-2014.

We also show in Fig. \ref{fig:average}(top, right) the relative standard deviation (RSD) of the growth rates distribution, defined as the standard deviation of the growth rates of all cities inside each departments divided by the average. This variance displays a decreasing behavior with time which implies that the aggregate growth rate for each departments is always closer to the value of the growth rate for a typical municipality inside the region.

At this stage, these results show that it is difficult to consider municipalities as isolated entities and that their growth rate is affected by neighbors. In particular, we see here that for municipalities belonging to the same urban area a global homogeneization trend where growth rate fluctuations disappear in time and in space.

\subsection*{Quantifying the coupling accessibility-growth rate}

We now focus on the coupling between accessibility and population growth rate. We considered different measures of accessibility but we will show here the results obtained for the accessibility used by Hansen \cite{Hansen:1959} which is defined below. Our results are however valid for other measures such as the inverse time to reach the closest train station or the inverse time to reach the center of Paris for example (see the SM, section 4 for details and discussions about various measures). The Hansen accessibility measure integrates the coupling between the infrastructure and the land-use component in a single expression which is expressed as \cite{Hansen:1959}
\begin{equation}
\label{eq:Hansen_1}
A_i = \sum_j \frac{S_j}{T_{i,j}^\tau}
\end{equation}                   
where the sum is taken over all areas $j$ that can be reached from the $i$-th area. Such an expression takes into account the land-use component $S_i$ which characterizes the activity of the area (for instance the population or the number of jobs), and the transportation component $T_{i,j}$ which is the travel distance between areas $i$ and $j$. The exponent $\tau$ weights how much the travel times between the areas impact on accessibility and is taken here equal to one. We have used this Hansen potential accessibility because it is one of the most used in the modern literature, especially in quantitative geography studies \cite{Kotavaara:2011, Koopmans:2012}, and because has been found to be the one with the largest explanatory power \cite{Song:1996}. Instead of performing a regression analysis as it is usually done in most studies (for example \cite{Koopmans:2012,Duranton:2012, Garcia-Lopez:2015, Kotavaara:2011, Mayer:2015}), we will exhibit directly the effect of accessibility on growth rates. We first study the impact of accessibility on growth rates and we observe a very weak dependence (see SM, Fig.~S3), showing that there is virtually no impact. Such negative result is actually in line with most studies where only a very weak effect was found, in contrast with the impressive results of Hansen \cite{Hansen:1959}. As expected from the previous section, growth rate fluctuations are usually small and we don't expect very important effects. In particular, in the case of the Paris urban area, we can understand the main reason behind the failure of accessibility to account for growth rates (see SM, Fig.~S4 for more details and maps of growth rate and accessibility): most accessibility measures considered are essentially related to centrality, i.e. how close the municipality is close to the center (Paris here) and to denser areas. Growth rates do not have however this structure at all: on the contrary, we observe in the recent history, an inverted structure where further municipalities have a larger growth rate, which signals a suburbanization of the area.

\begin{figure*}[ht!]
    \includegraphics[scale=0.2]{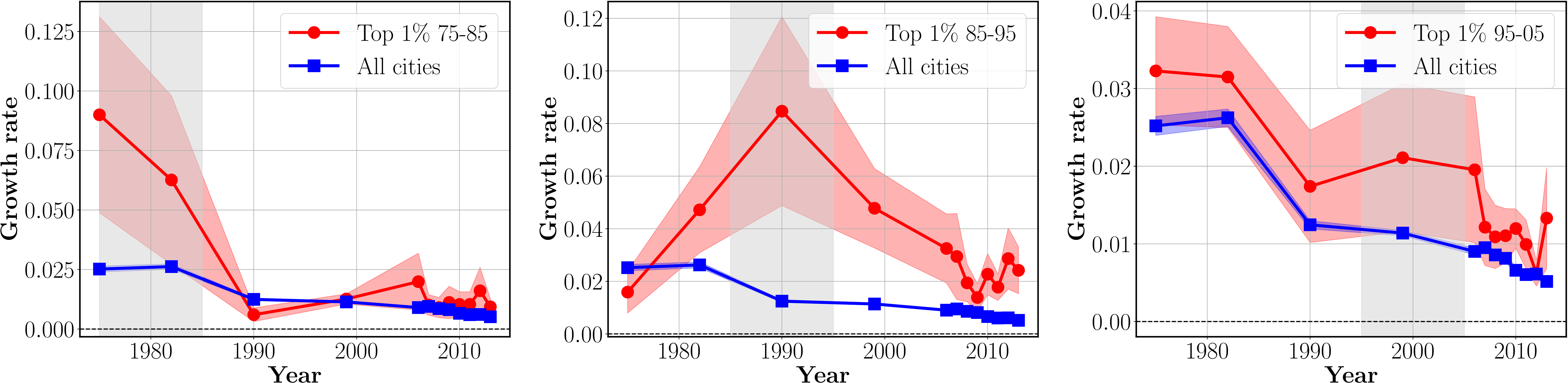}
\caption{Growth rate and accessibility variations. We first rank cities according to their  variation of the Hansen accessibility (in the period 75-85, for the leftmost, in the period 1985-1995 for the central figure, and in the period 1995-2005 for rightmost panel), and measure the average of the growth rates for the cities in the top $1\%$ (in red) and for all cities (in blue).  As we can see in the leftmost figure, the cities which experienced a large accessibility variation in the period 1975-1985, have a significantly larger average growth rate in the same period. The same effect is visible for the time period 1985-1995 in the central panel, while for the last time period 1995-2005, we observe a smaller value of this difference in growth rates.}
\label{fig:A2-average} 
\end{figure*}

These negative results lead us to consider cities that experienced a \emph{variation} of accessibility. This seems reasonable as an improvment of transportation mode in a given area can indeed trigger a wave of newcomers. Empirically, if we first consider all cities together, we don't observe any significative trend. We then focused on municipalities with the largest accessibility variation such as the top $1\%$ of cities who display the largest accessibility increase in a given period of time. In Fig.~\ref{fig:A2-average}, we show the growth rate for the top $1\%$ and for all cities for different time periods. We observe that cities with larger variations of Hansen accessibility display indeed a significantly larger average growth rate compared to the average (see Figs.~S5, S6 in the SM for a discussion with other accessibility measures). This effect is in particular present for the periods 1975-1985 and 1985-1995, while for the period 1995-2005 the effect is less significant.  
This impact identified in Fig.~\ref{fig:A2-average} is stable and significant only if we consider the municipalities that witnessed the largest accessibility variations (the cities in the top $1-2\%$). In the SM (Fig.~S7), we show
that the difference between the growth rates of all cities and the selected subset of cities with a large variations becomes negligible as soon as more than $2\%$ of the cities are considered. The effect observed here seems therefore to be relevant only for a small fractions of cities. This is probably related to the fact that in the periods considered here, the transportation networks did not evolve dramatically: for most of these periods, only $5\%$ to $10\%$ of new sections are added to the network (see Fig.~S8 in the SM). In addition, we observe on  Fig.~\ref{fig:A2-average} a rapid decrease of the difference between the growth rates for the two groups: for the period 1975-1985, we observe that after 1990 the growth rates are simlar, and for the period 1985-1995, there are no significative differences after 2010 approximately. These results therefore point to two important conclusions:
\begin{enumerate}
\item{} First, instead of accessibility, it is the \emph{accessibility variation} that acts as a control parameter on the population growth rates.
\item{} Second, the impact of accessibility variation is limited in time and growth rates rapidly converge back to the average value for all cities in the considered area.
\end{enumerate}
These two remarks are obviously important pieces of the puzzle that we will use for constructing a simplified model of this effect.

\section*{Modeling the impact of accessibility}


In order to gain a further understanding and quantitative insights about the coupling between accessibility and growth rate, we introduce here a simple model. We start from the general diffusion equation with noise which reads
\begin{align}
\frac{d P_i}{d t} = \eta_i P_i +\sum_jJ_{ji}P_j-J_{ij}P_i
\end{align}
This equation which was introduced in the context of wealth dynamics 
\cite{Bouchaud:2000}
has a natural interpretation in the case of cities \cite{Barthelemy:2016}. The first term corresponds to the Gibrat model \cite{Gibrat:1931} and describes the 
stochastic growth of population (birth-death processes and other exogenous processes) and the random variable $\eta_i$ is assumed to be a  Gaussian noise, with average $m$ and variance $2 \sigma^2$. 
The other terms describe migration of individuals from one city to another: $J_{ij}$ is the migration rate from city $j$ to city $i$. It has been shown that this equation provides a regularization of the Gibrat model and a natural explanation of Zipf's law, at least in the mean-field version where $J_{ij}=J$ for all $i$ and $j$ \cite{Bouchaud:2000,Barthelemy:2016}. In the Gibrat case ($J_{ij}=0$) the growth rate does not depend on the population and fluctuates around the average value
\begin{equation}
\langle g \rangle := \frac{1}{\langle P\rangle}\frac{d \langle P \rangle }{d t } = m + \sigma^2 
\end{equation}
where $\langle\cdot\rangle$ denotes the average over the noise $\eta$.

\begin{figure}
  \includegraphics[scale=0.5]{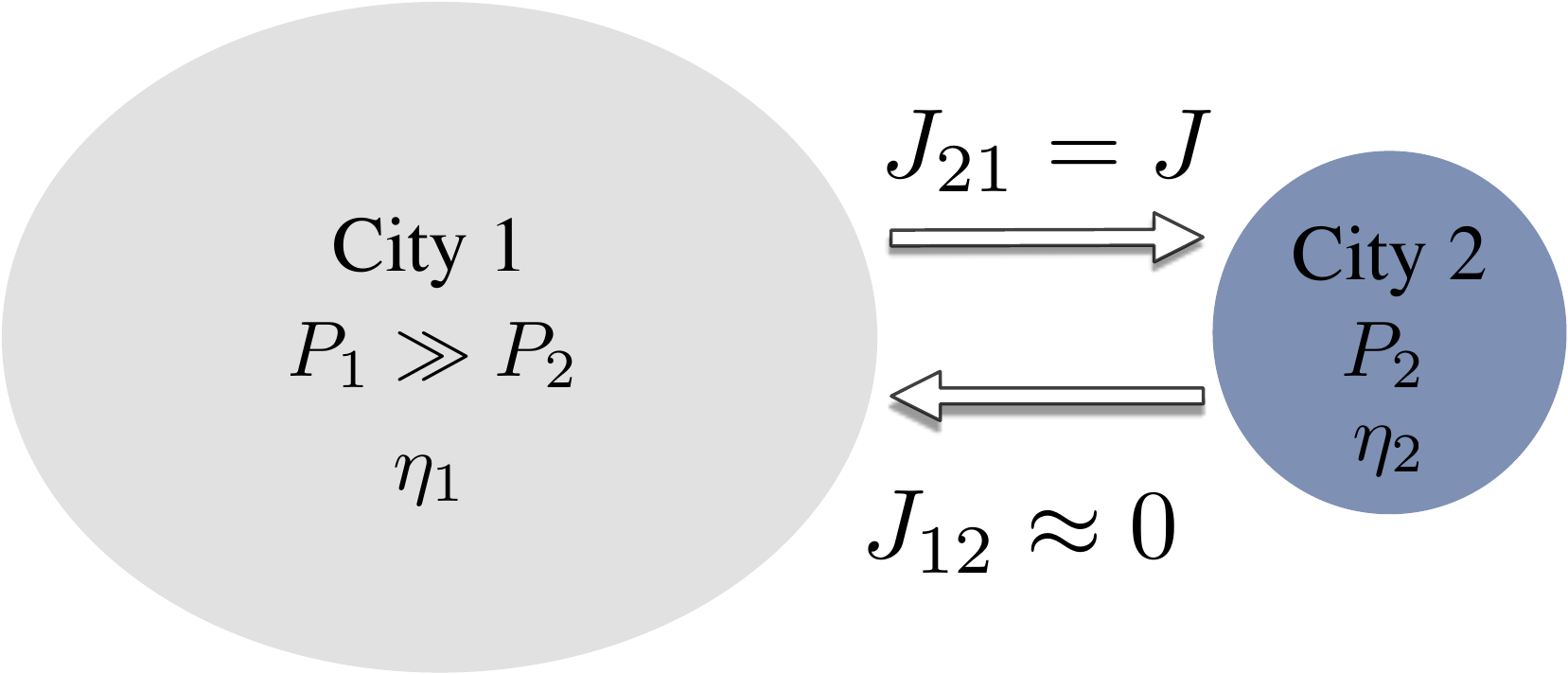}
\caption{Schematic representation of the model. The city 1 is much larger than city 2 (and could represent the `rest of the world' for city 2. The internal growth for each city is described by a random gaussian noise $\eta_i$ ($i=1,2$) with the same average and variance. There is a migration flow $J$ from 1 to 2 and we neglect the reverse flow from city 2 to city 1.}  
\label{fig:model}
\end{figure}

We now introduce a minimal model for the impact on population growth of increasing accessibility which consists of two cities, 1 and 2. We have in mind a large city 1 connected to a small peripheral city 2 with $P_2 \ll P_1$ (City 1 can in fact be considered as the whole world outside city 2, see Fig.~\ref{fig:model}). 
We also assume that there is a migration from city 1 to city 2 that is described by the rate $J$ and we neglect the counterflow from 2 to 1. The evolution of populations 1 and 2 is then given in this framework by
\begin{align}
\label{eq-2cities}
\frac{d P_1}{d t} &= \eta_1 P_1   \\
\frac{d P_2}{d t} &= \eta_2 P_2 + J P_1 \nonumber
\end{align}
where we assume that both $\eta_1$ and $\eta_2$ have the same average $m$ and variance
$2 \sigma^2$. In the case where the cities are disconnected ($J = 0 $), they grow with the same natural rate
\begin{equation}
g_1 = g_2 = \frac{d \langle P_{1(2)}\rangle}{\langle P_{1(2)}\rangle d t} = m + \sigma^2.
\end{equation}
When there is a flow $J>0$ from city 1 to city 2, the formal solutions of equations \eqref{eq-2cities} are
\begin{align}
\nonumber
P_1(t) &= P_1(0) \left( e^{\int_0^t \eta_1 (t')}  \right) \\
P_2(t) &= P_2(0)e^{\int_0^t \eta_2 (t')} + \int_0^t J(t') P_1(t') e^{\int_{t'}^{t} \eta_2 (t'')}. 
\end{align}
where we consider the general case where $J$ depends on time, and where we used the fact that the equation for $P_1$ is decoupled. For understanding the impact of accessibility variations on the growth rate, we consider the following simple scenario for the time-varying migration rate $J$. The city 2 is coupled to city 1 at time $t=t_0$ and we assume that the coupling lasts a finite duration $\delta t$: 
\begin{align}
J(t) &= 0 \ \ \text{for}\ \  t < t_0  \\
J(t) &= J \ \ \text{for}\ \  t_0 < t < t_0 + \delta t \nonumber \\
J(t) &= 0 \ \ \text{for}\ \  t > t_0 + \delta t \nonumber 
\end{align}
For this simple scenario, we can compute the growth rate $g_2$ of city 2 and find in the limit where the number of newcomers in city 2 is much less than the population $P_2$ (see SM, section 7 for more details)
\begin{align}
g_2 &= m + \sigma^2 \ \ \text{for}\ \  t < t_0  \\
  g_2 &= (m + \sigma^2) + J \ \frac{\langle P_1(t) \rangle}{\langle P_2(t) \rangle}  \ \ \text{for}\ \  t_0 < t < t_0 + \delta t \nonumber \\
g_2 &= m + \sigma^2 \ \ \text{for}\ \  t > t_0 + \delta t \nonumber .
\end{align}
where $\langle P_1(t)\rangle/\langle P_2(t)\rangle\approx P_1(0)/P_2(0)$ when the number of newcomers in city 2 is much less than the population $P_2$ (see SM7 for more details). This simple model thus predicts an increase for growth rates that is linear in $J$, and inversely proportional to the population of city 2. The growth rate is thus larger during the migration period and is back to its uncoupled value afterwards, in agreement with our empirical observations and which justifies this finite duration $\delta t$.

\begin{figure*}[htb!]
  \includegraphics[scale=0.2]{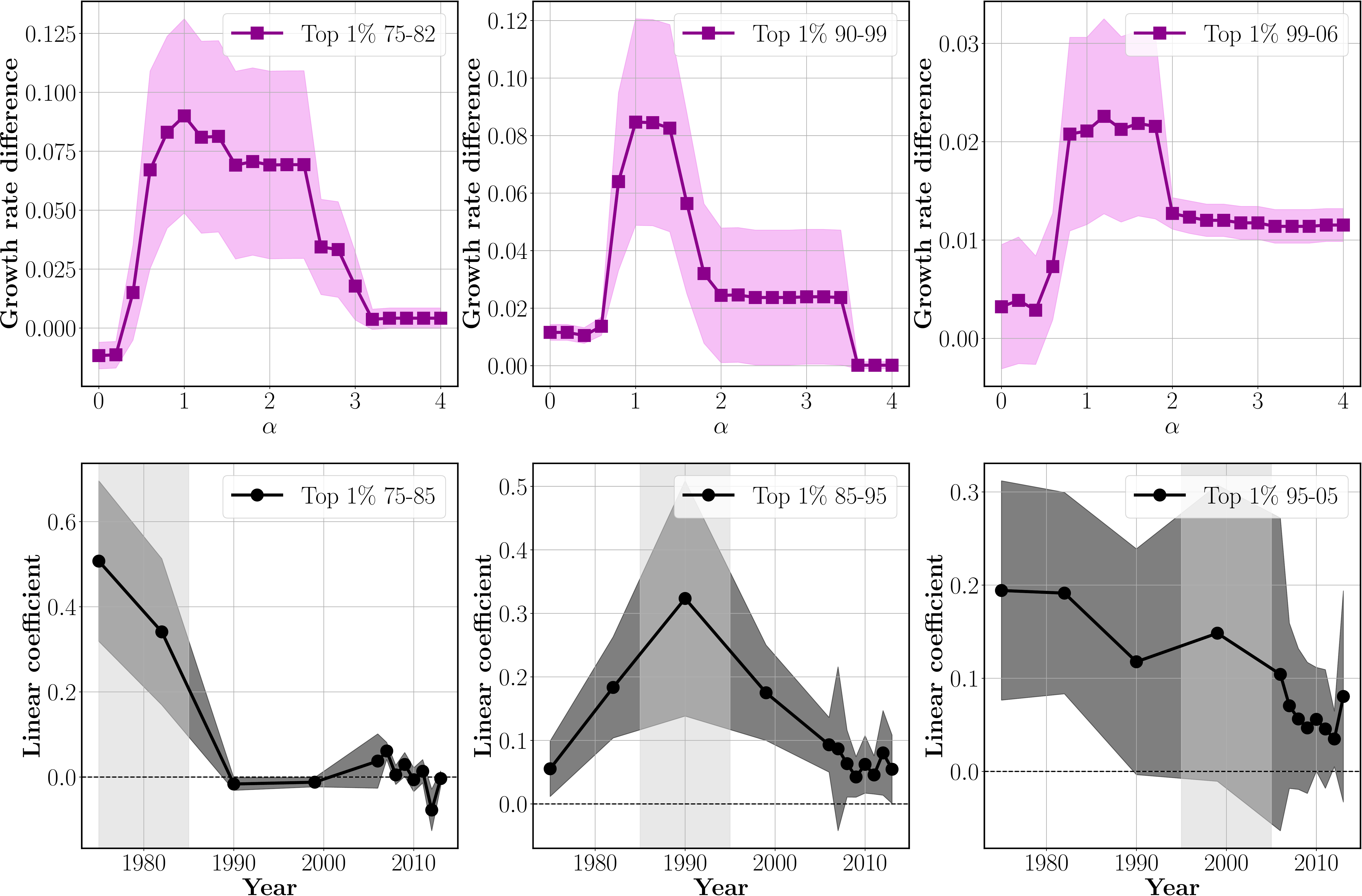}
\caption{(Top) Varying $\alpha$. We plot the difference between the growth rate of cities which experiences some change and the rest. Here the cities are ranked according to their value of $\Delta A P^{\alpha-1}$. We see that the most significant difference between the most affected cities and the rest is observed for $\alpha \simeq 1$. (Bottom) Variations of the growth rate with the Hansen accessibility (for the top $1\%$ of cities ranked according to $\Delta A$). We plot here the linear coefficient $\kappa\equiv \Delta g/\Delta A$ (Eq.~\ref{eq:linear})  that describes the impact of accessibility on growth. This linear coefficient is significantly different from zero for the growth rates corresponding to the periods of observed accessibility variations. These results are significant for the periods 1975-1985 and 1985-1995 only.}  
\label{fig:A2-alpha}
\end{figure*}

In order to connect this model to real-world data, we have to identify the migration rate $J$. From our empirical results, it seems natural to identify $J$ with the accessibility variations $\Delta A$ and not with the accessibility $A$ of the area considered.  We also assume that the impact of a given accessibility variation is larger for a larger city which implies that $J$ is an increasing function of the population $P$, and we will assume a simple power law form $P^\alpha$. This leads us to the main assumption of our model that consists in writing the coupling $J$ as
\begin{align}
J\propto P^{\alpha}\Delta A
\end{align}
where $P$ is the average population of city 2 and $\Delta A$ the accessibility variation experienced by this city. With this expression, the growth rate of city 2 is given by
\begin{align}
g_2 &= m + \sigma^2 \ \ \text{for}\ \  t < t_0  \\
g_2 &= (m + \sigma^2) + \kappa P^{\alpha-1} \Delta A  \ \ \text{for}\ \  t_0 < t < t_0 + \delta t \nonumber \\
g_2 &= m + \sigma^2 \ \ \text{for}\ \  t > t_0 + \delta t \nonumber .
\end{align}
where we see that the impact of the growth rate variation scales as $ \kappa P^{\alpha-1} \Delta A$ (we assumed here that $\langle P_1\rangle$ is varying very slowly over the time scale $\delta t$ and we integrate it in the constant $\kappa$). Using this result, we can determine the value of $\alpha$ from empirical data. Indeed, for fixed $\alpha$, the growth rate variation of a city is proportional to $P^{\alpha-1} \Delta A$ and we look at the value of $\alpha$ which gives the largest difference between the top $1\%$ and the rest. We thus assume here that the best value of $\alpha$ is the one that maximizes the impact of the accessibility variation on the system. We thus select cities with the largest value 
$ \kappa P^{\alpha-1} \Delta A$ and measure the difference of their growth rate with respect to the average. We show the results in Fig.~\ref{fig:A2-alpha}(top) which demonstrate that it is in the region $\alpha \simeq 1$ where we observe the most significant differences among cities with the largest migration inflow and the rest. This result also confirms a posteriori our empirical analysis of growth rates versus accessibility variations. 

This model -- together with the empirical result $\alpha\approx 1$ -- thus provides a basic mechanism for the coupling between accessibility and growth rate variations, and predicts that the growth rate variation depends linearly with the accessibility variations
\begin{equation}
  \Delta g\simeq \kappa \Delta A .
  \label{eq:linear}
\end{equation}
where $\Delta g$ is the difference between the growth rates after and before the accessibility change. As we have seen however, a significant effect seems to be quantifiable only for the few $0.5$ or $1\%$ of cities who experienced the largest variation of accessibility. We therefore focus on this subset of cities which experienced a significative change in accessibility, and test if such a linear dependence can indeed be observable. We plot in Fig.~\ref{fig:A2-alpha}(bottom) the linear coefficient of the regression for the different periods. As expected, we observe that this linear coefficient is significantly different from zero for the periods of observed accessibility variations (here this effect is significative only for the periods 1975-1985 and 1985-1995, see SM, section 8 for more details), and we observe that $\kappa$ is of the order $10^{-1}$ with values at most equal to  $0.5$ in our data.

\section*{Discussion}

Accessibility measures are expected to be a powerful tool to build explanatory variables that have a large predictive power on growth. There is a large literature in quantitative geography and spatial econometrics that points in this direction, making accessibility as an extremely useful tool to assess the impact of transportation modes on growth.  However, we have shown here that the relevant variable seems to be the variation in time of the accessibility (as already suggested in \cite{Mayer:2015}). Also, it seems that the effect of such a variation decays in time (and space) and that cities recover relatively quickly a population growth similar to the average of the corresponding region. These different elements led us to propose a simple model where the important ingredient is the interurban flow that has a limited lifetime and which is proportional to the accessibility variation. This model is a first step towards the modelling of the accessibility-growth rate coupling, and provides a framework that can be built upon. It will allow to go beyond regression analysis, and eventually to help planners for identifiying critical factors for the evolution of cities.


{\bf Acknowledgements}. VV thanks the IPhT for its hospitality.

\section*{Bibliography}



\clearpage

\newpage

\newpage

\renewcommand\thefigure{\thesection.\arabic{figure}}    
\setcounter{figure}{0}

\renewcommand{\thefigure}{S\arabic{figure}}

\section{Supplementary Material}

\subsection{Dataset description}

We use data freely available from the National Institute of Statistics and
Economic Studies (INSEE) \cite{Insee}. The dataset contains the
population of each municipality in France for the years 1968, 1975,
1982, 1990, 1999, and all years from 2006 to 2014.

The number of municipalities is not the same every year, due to
merging and separation of administrative units. On the contrary, it
fluctuates from a minimum of 35,891 municipalities in 1982, to a
maximum of 37,727 in 1968.

Here we focus at first on the 35,513 municipalities that are present
in the INSEE list for all the years, and more specifically, we will
concentrate on a subset of 4,457 municipality belonging to the 10
largest urban areas in France.

In order to understand how the evolution of population of french
municipalities depends on the geographical location, we geo-localized
each individual municipalities assigning to it longitude and latitude
of its centroid. We refer to the position of municipality $i$ by the
vector $\vec{x}_i$.

\begin{figure}[ht!]
  \includegraphics[scale=0.6]{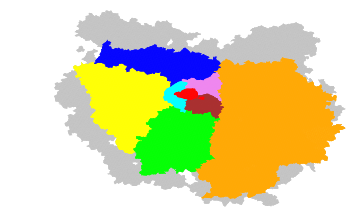} 
\caption{Map of Ile-de-France and their departments divisions: Seine-et-Marne(77), Yveline(78), Essone(91), Haute-de-Seine(92), Seine-Saint-Denis(93), Val-de-Marne(94) and Val-d'Oise(95).}
\label{fig:map}
\end{figure}

In particular, we will consider the region Ile-de-France that contains and surrounds Paris and
which is shown in Fig.~\ref{fig:map}.

\subsection{Growth rates: fluctuations and spatial disparities}

We define the growth rate at time $t$, $g_i(t)$  as
\begin{equation}
\label{eq:gr_definition}
g_i (t) = \frac{P_i(t+\delta t) - P_i(t)}{P_i(t) \delta t}.
\end{equation}

\begin{figure}
    \includegraphics[scale=0.3]{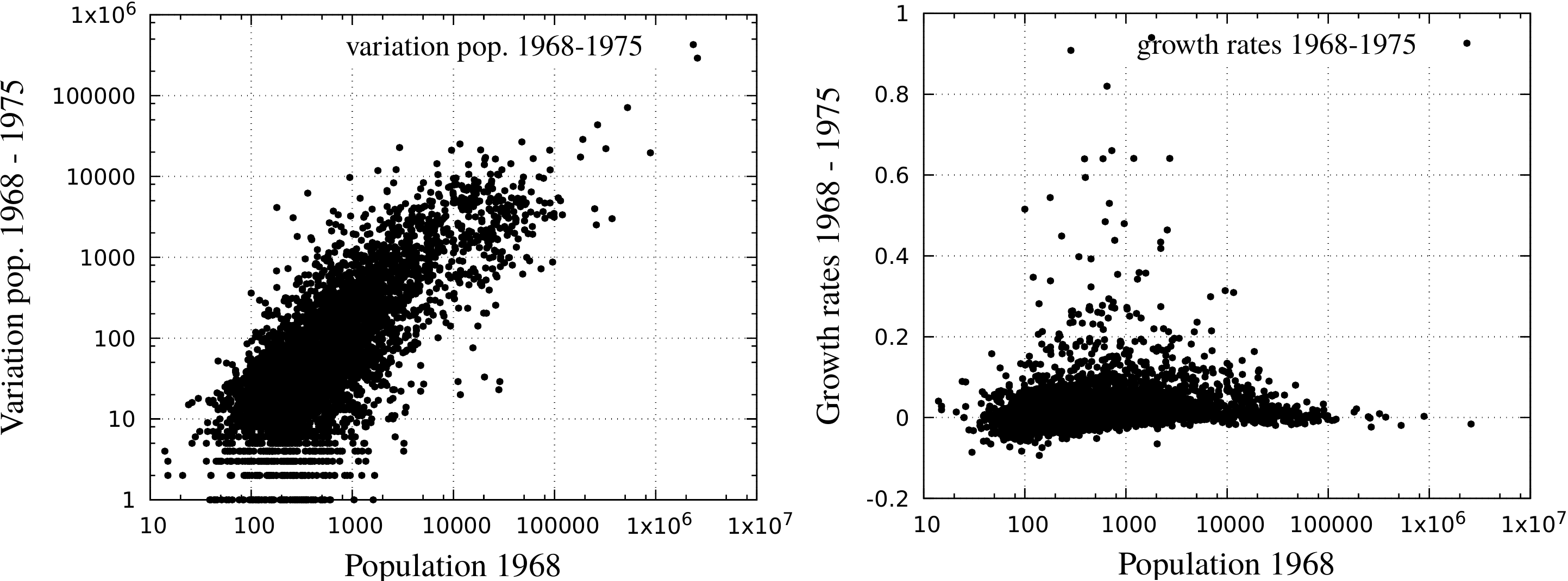}
\caption{Variations of population and growth rates. We show here the dependence of the
  variations of population $|P_i(t+\delta t) - P_i(t)|$ (left) and
  growth rates \eqref{eq:gr_definition}, (right)  versus the population $P_i(t)$, for the set of 4,457cities in our
 dataset, in the period 1968-1975. While the variations of populations
 grow linearly with the $P_i(t)$, the growth rates are essentially 
 independent on the population.}
\label{fig:gr}
\end{figure}

The growth rates normalized as in \eqref{eq:gr_definition} seem to be
independent from the population as we can observe in Fig. \ref{fig:gr}.

\subsection{The failure of accessibility}

\begin{figure}
  \includegraphics[scale=0.6]{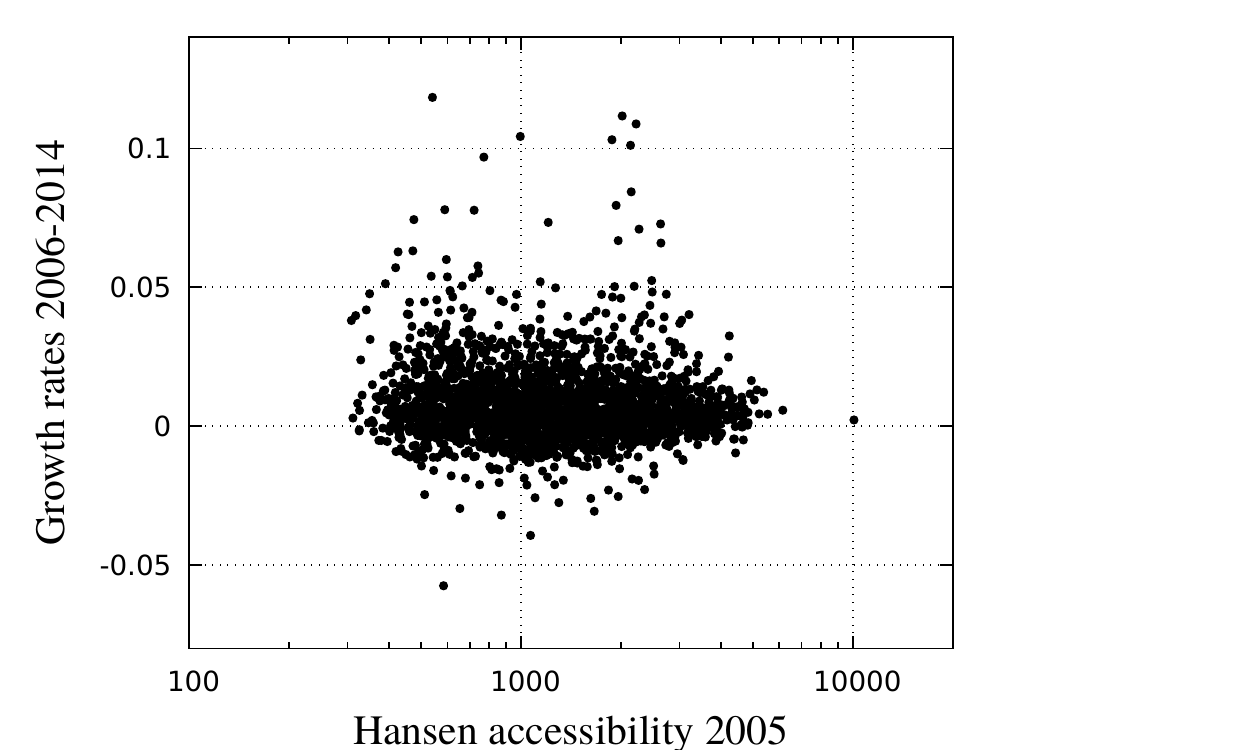}
\caption{ \textbf{ Population growth rates versus Hansen
    accessibility}.  A very weak dependence is found, in sharp
  constrast with results shown in the seminal paper
  \cite{Hansen:1959}. The dependence here can be fitted with a power
  law or with a linear relation, but results with very low
  values of $\chi^2$ are obtained, meaning that the fit is very
  dependent on the growth rates of a very small number of
  municipalities, making the result not statistically significant.}
\label{fig:A2-GR}
\end{figure}

In Fig. \ref{fig:A2-GR}, we show the impact of Hansen potantial
accessibility (Eq.~ \ref{eq:Hansen_1}) on population growth rates,
for the 2301 municipalities in the Paris urban area.
Such negative result is in line with the discussion in the literature
review. In fact, in most studies, a very weak effect of accessibility
is found in recent years. From the very impressive result of Hansen
\cite{Hansen:1959}, recent studies find that accessibility impact is
much more weak, and in particular is significant in early years of
studies \cite{Kotavaara:2011,Koopmans:2012}.

\begin{figure}
  \includegraphics[scale=0.58]{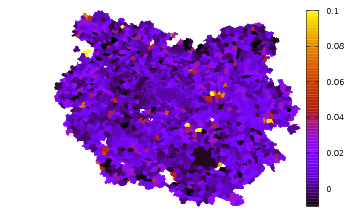}
  \includegraphics[scale=0.18]{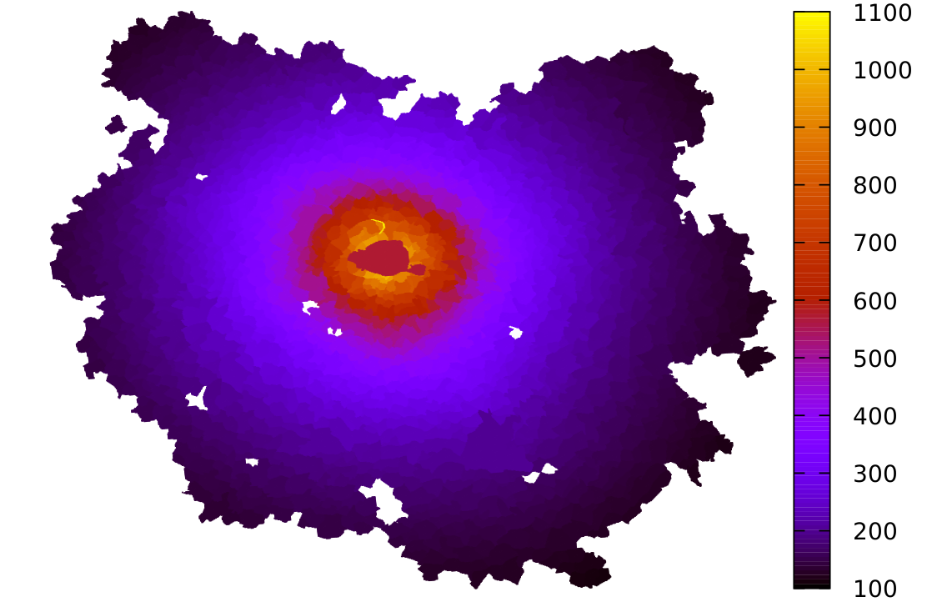}
\caption{(Top) Growth rates in Ile-de-France for the period
    2006-2014. (Bottom) Hansen accessibility in
    the Ile-de-France region for the same period 2006-2014.}
\label{fig:map2}
\end{figure}

In Fig. \ref{fig:map2} we show the maps for accessibility and growth
rates in the Paris urban area, for the most recent available
years. Here we see what could be the main reason behind the failure of
accessibility to account for growth rates. Potential accessibility --
like all accessibility measures considered, is essentialy a central
measure, i.e. municipalities close to the center (Paris) and closer to
denser areas, are the one with largest value of accessiblity. Growth
rates however do not have at all this structure. On the contrary,
often in recent history an inverse structure is exhibited
(sub-urbanization).

\subsection{Accessibility measures}

Accessibility is a concept used in a large number of studies in
Economics, Geography and urban planning, in order to capture the
potential of a given area to grow.  Several measures of accessibility
have been proposed in the literature, and the different approaches are
reviewed in \cite{Ingram:1971, Jones:1981, Koenig:1980,
  Handy:1997,Geurs:2004}.  In most of the accessibility measures that
we are going to discuss here, an essential ingredient is the time
needed to travel from one location to another. The growth of the
public transportation system induces a decrease of these travel times
and in general improves the accessibility at the global level and
locally for areas where new stations and new lines are built. For this
reason, assessing accessibility improvements following the expansion
of the public transportation network is a key element for estimating
benefits of public investments in urban planning.  Accessibility
measures can be classified into five general families :
\begin{itemize}
\item Local accessibility measures
\item Potential accessibility measures
\item Polar accessibility measures
\item Contour accessibility measures
\item Network accessibility measures
\end{itemize}

The first family, is composed by measures which define the
accessibility of an area in terms of local quantities only, such as
local properties of the infrastructure network. This can be the number
of train stations or road density in the area \cite{Duranton:2012}, or
the distance to the closest train station \cite{Garcia-Lopez:2015}. It
can also be related to densities of different activities taking place
in the area (eg. density of jobs or shopping centers). An important
limitation of these measures is that their refer exclusively to a
property of the transportation network (the infrastructure component)
or exclusively to some activities (the land-use component) without
taking into account their interaction. An additional limitation of
these local measures is that correlations and interactions between
different areas are ignored. We can then observe with these measures
non realistic situations such as low accessibility areas contiguous to
high accessbility ones (while we would expect a smoother decrease).

The second family comprises measures that integrate the coupling
between the infrastructure and the land-use component in a single
expression. They are usually defined in terms of a local density
quantifying the size of a given activity in the area, $S_i$ (for
instance the population in the area or the number of jobs), and in
terms of the travel distance between areas, $T_{i,j}$ . The
accessibility of the $i$-th area is then expressed, as in reference
\cite{Hansen:1959}, as
                                
\begin{equation}
\label{eq:Hansen_1}
A_i = \sum_j \frac{S_j}{T_{i,j}^\tau}
\end{equation}                               
where the sum is taken over all areas that can be reached from the
$i$-th area. Such an expression takes into account the land-use
component ($S_i$) and the transportation component ($T_{i,j}$) in the
same measure. The exponent $\tau$ weights how much the travel times between
the areas impact on accessibility (and we will assume here $\tau=1$).  In several application (see
\cite{Kotavaara:2011, Koopmans:2012}), a self-potential term is added
to equation \eqref{eq:Hansen_1} in order to include the internal
movements inside area $i$. \eqref{eq:Hansen_1} is then
generalized to
\begin{equation}
\label{eq:Hansen_2}
A_i = \frac{S_i}{T_{i,i}^k} + \sum_j \frac{S_j}{T_{i,j}^\tau}
\end{equation}  
where $T_{i,i}$ is a characteristic time of movements inside the
$i$-th area.  The empirical justification that historically leads to
such a definition of accessibility finds its orgin in the observation
of human mobility patterns. In a seminal paper \cite{Zipf:1946} it has
been observed that the number of individuals $P_{i,j}$ that move
between locations $i$ and $j$ per unit time usually follows the
so-called gravity law given by
\begin{equation}
\label{eq:Zipf_gravity}
P_{i,j} = \frac{P_i P_j}{f (T_{i,j})}
\end{equation}                            
where $P_i$ ($P_j$) is the population of the $i$-th ($j$-th) location
and the function $f(T)$ quantifies the impact of travel time on the
population flows. Such a function is usually assumed to be a power law
which leads to expressions found in Eqs. \eqref{eq:Hansen_1} and
\eqref{eq:Hansen_2} for defining the accessibility.  The law equation
\eqref{eq:Zipf_gravity} has however been criticized in the recent
literature \cite{Simini:2012} because of the lack of a clear
theoretical justification and discrepancies when compared to recent
empirical observations. We mention briefly here that thanks to the
recent ICT revolution, huge amounts of data on human mobility can be
easily accessed and new empirical properties on mobility patterns can
be found and discussed (see for example the review \cite{Barbosa:2018}). Despite their opaque meaning, and the lack of a
theoretical derivation, measures of accessibility based on the gravity
assumption are probably the most commonly used in econometric
regression analysis (see for instance \cite{Kotavaara:2011,
  Koopmans:2012}), in order to assess the impact of various
transportation modes on urban development.

Measures in the third family consider the potential of an area
computed with respect to only specific location, usually the center of
the corresponding urban area. In \cite{Mayer:2015}, for instance,
the time distance from Paris has been used to quantify the
accessibility of the different municipalities in this urban area. This
type of measures depends only on the transport component of
accessibility, with no reference to the land-use one, but are not
local, since the accessibility of a given area depends on the global
properties of the transport network.

The fourth family of measures compute the density of land-use
activities (or opportunities) that can be reached from a given area,
within a given travel time (or distance or cost). When time is used as
a constraint, they are often referred as isochrone measures. The
opportunities could for example be the number of jobs or different
services that are present in the area that can be reached from the
location within the isochrone time \cite{Wachs:1973}, or in some cases
even simply the surface of the area than can be covered. Measures of
this type are frequently used by practitioners, mostly because they
are intuitive, but might present problems due to the arbitrary
selection of the isochrones of interest \cite{Geurs:2004}.

As already observed, in order to be computed, all the accessibility
measures considered so far need the calculation of travel times from
any location to any other one. We refer to this matrix of travel times
as the shortest path matrix. A fifth class of measures can then be
defined as the measures that compute some network-based quantities,
starting from this matrix.

\subsection{Growth rates: fluctuations and spatial disparities}

\begin{figure*}
  \begin{center}
    \includegraphics[scale=0.5]{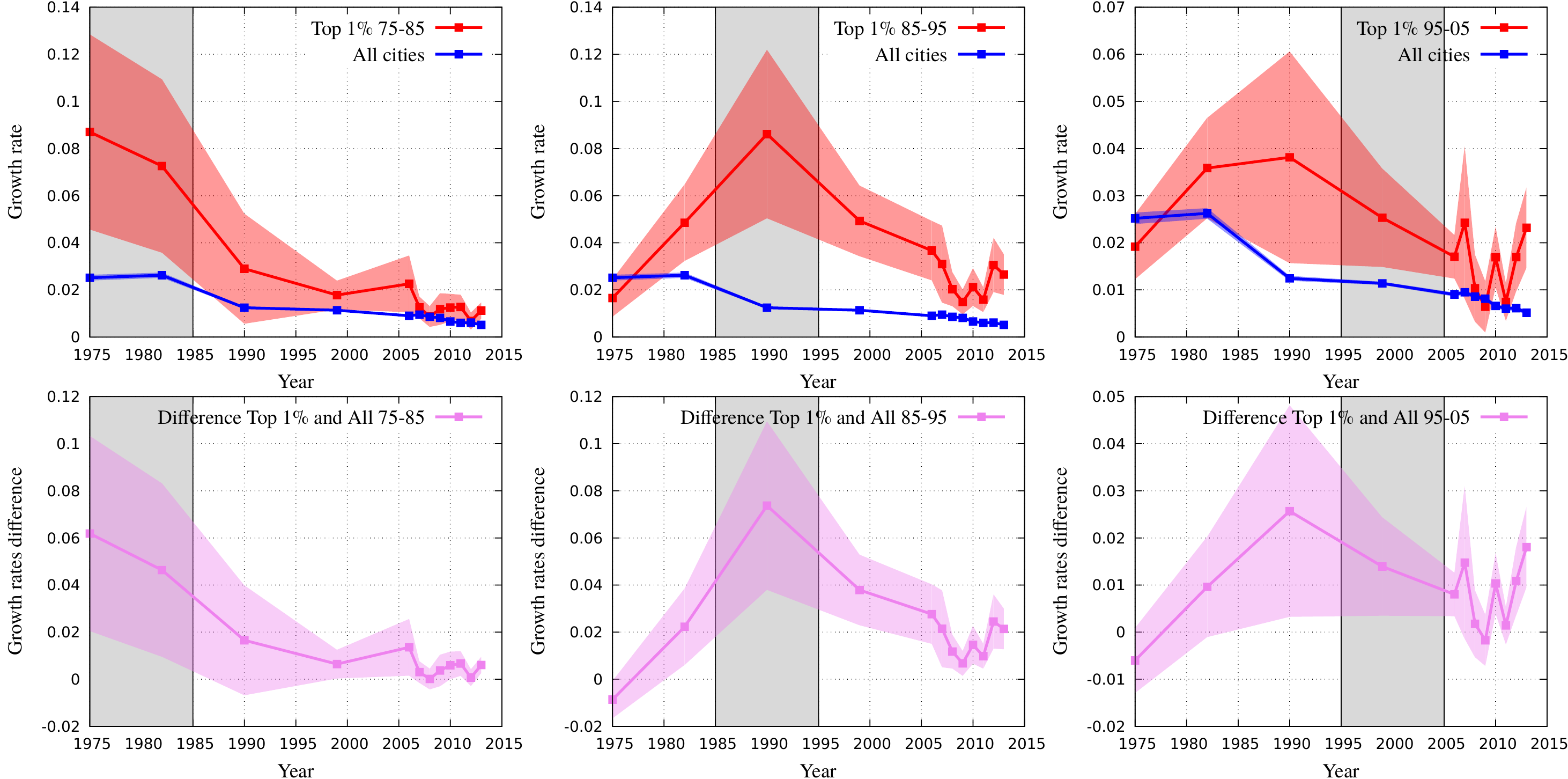}  
\caption{\textbf{Variations of accessibility: Inverse time to reach
    the closest station}. For these figures, we ranked the cities
  according to the variations of the inverse time to reach the closest
  station. For the leftmost panel, we used the variation of
  this accessibility in the period 75-85, and looked at the average of
  the growth rates for the cities in the top $1\%$ (in red, top) and
  for all cities (in blue, top). For the bottom panels, the difference between the average growth rate of the $1\%$
  and all cities is showed (in purple).  As we can see, the cities
  which experienced a large accessibility variation in the period
  75-85, have a significantly larger average growth rate than the
  rest, in the period in which this accessibility variation has been
  produced, but this difference decreases with time until it becomes not
  significative. }
\label{fig:A1-average}
\end{center}
\end{figure*}

\begin{figure*}
  \begin{center}
    \includegraphics[scale=0.5]{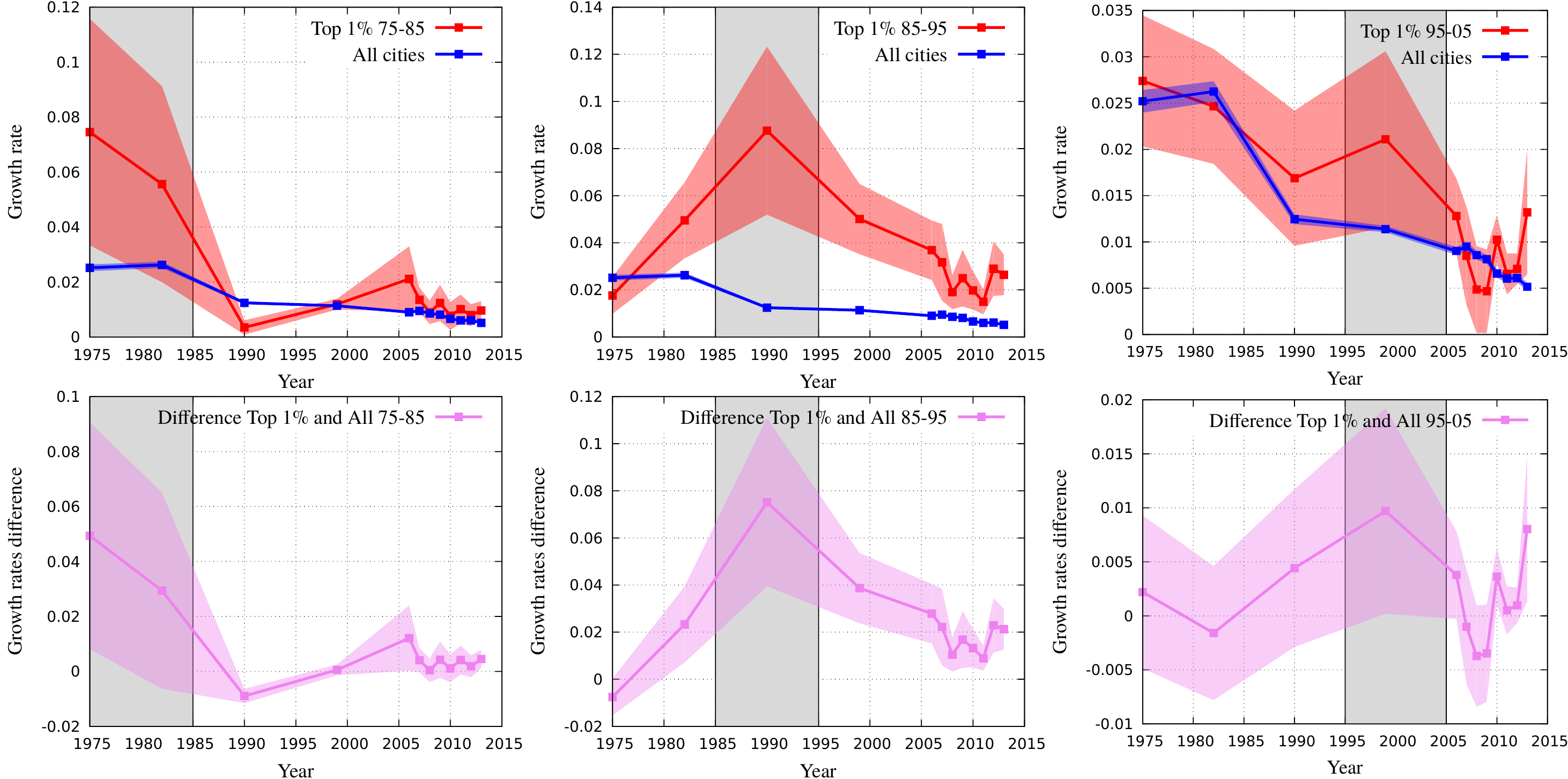}  
\caption{\textbf{Variations of accessibility: Inverse time to reach
    the center of Paris}. For these figures, we ranked the cities
  according to the variation of the inverse time to reach the center of
  Paris. For the leftmost panel, we used the variation of this
  accessibility in the period 75-85, and looked at the average of the
  growth rates for the cities in the top $1\%$ (in red, top) and for
  all cities (in blue, top). For the bottom panels, the difference between the average growth rate of the $1\%$
  and all cities is showed (in purple). As we can observe, cities
  which experienced a large accessibility variation in the period
  75-85, have a significantly larger average growth rate than the
  rest, in the period in which this accessibility variation has been
  produced, but the difference decreases in time until it becomes not significative.}
\label{fig:A3-average}
\end{center}
\end{figure*}

 In Fig.~\ref{fig:A1-average} and \ref{fig:A3-average} we show that the cities
 that have the larger variations of accessibility (the top 1\%) show a
 significantly larger average growth rates, with respect to the
 remaining cities. This effect is in particular evident for the
 periods 1975-1985 and 1985-1995, while for the period 1995-2005 the
 effect is less significant. 

\subsection{Average growth rates, the range of significance}

\begin{figure*}
  \includegraphics[scale=0.6]{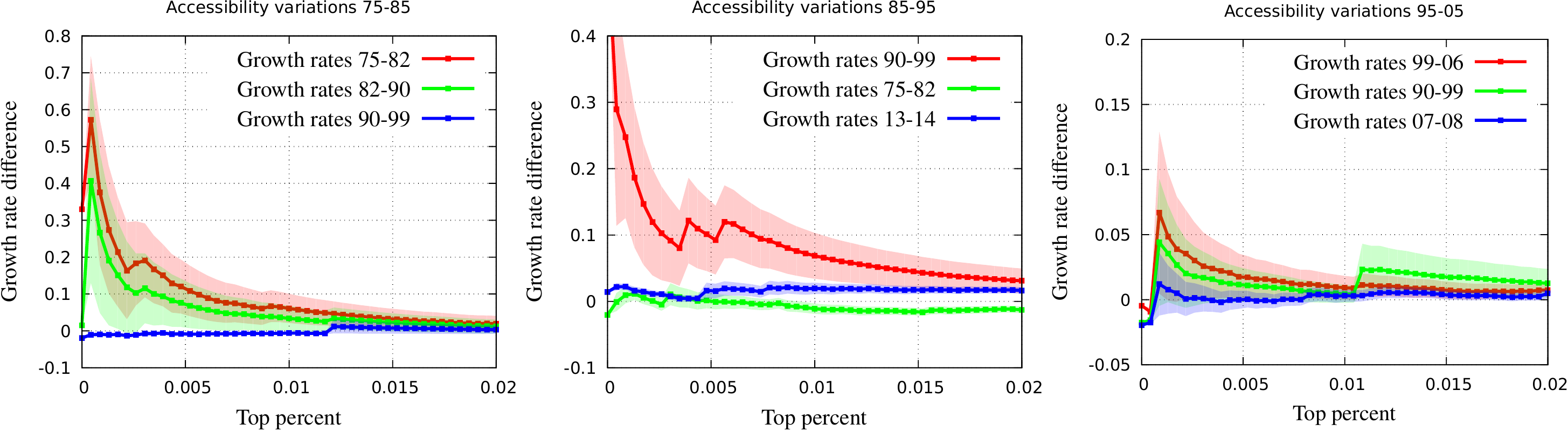}  
\caption{\textbf{Hansen accessibility - range of significance}.  For
  these figures, we ranked cities according to variations of
  Hansen accessibility (in the period 75-85, for the leftmost panel), and
  looked at the average of the growth rates for the cities in a given
  top percentile and for all cities. As we can observe, for the period 75-85 in the leftmost figure,
  for the growth rates of the 75-82 and 82-90, cities
  which experienced a large accessibility variations display also a significantly larger average growth rate than the rest, but this
  effect is significant only when the top $1\%$ is considered }
\label{fig:A2-differences}
\end{figure*}

In Figure \ref{fig:A2-differences} we show how ranking cities
according to the variations in the Hansen accessibility measures,
differences in average between the most affected cities and the rest
becomes negligible, as soon as more then $2\%$ of the cities are
considered.

\begin{figure}
  \begin{center}
    \includegraphics[scale=0.25]{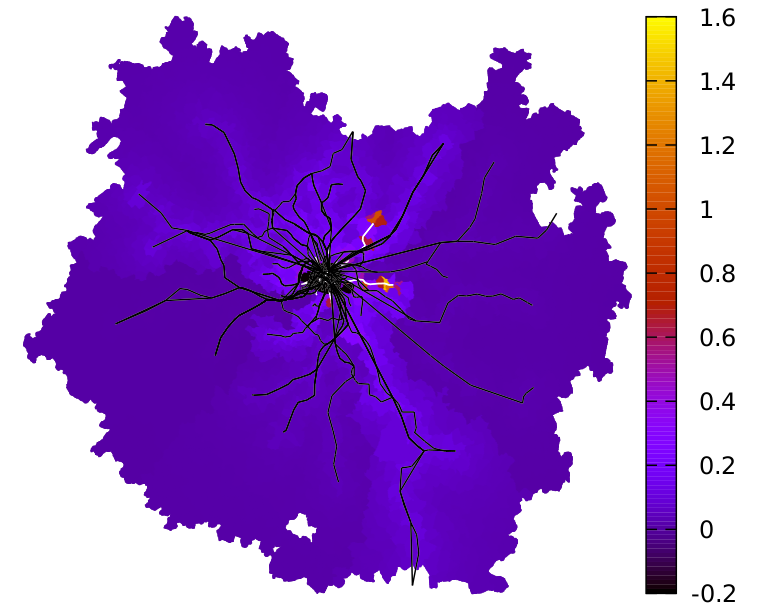}  
  \caption{\textbf{Variations of accessibility: Inverse time to reach
      the center of Paris 1975-1985}. We show here the map of the
    urban area of Paris where each municipality has a color depending
    on its accessibility variation. The black lines represent the
    transportation network in 1975, and in
    white lines the part of the network added in 1985.}
\label{fig:DA3}
\end{center}
\end{figure}

The effect seems to be relevant for a small fractions of cities only
and to the fact that in the periods considered, the
transportation networks do not evolve dramatically. In fact, from a
time snapshot to the other, only $5\%$ to $10\%$ of new sections are added
to the network. In figure \ref{fig:DA3} we show on a map of the urban
area of Paris, the few changes happened on the transportation network
from 1975 to 1985, and the municipality who experienced a change in
accessibility, measured in terms of inverse time to reach the center
of Paris.

\subsection{Modelling the coupling growth rates-accessibility variation}

The evolution of the population of a single city subjected to random
fluctuations can be written as
\begin{align}
  \label{eq-1city}
\frac{d P}{d t} &= \eta P
\end{align}
This simple evolution equation with multiplicative noise was discussed
by Gibrat \cite{Gibrat:1931}  and revised later by Gabaix \cite{Gabaix}.
This dynamics describes the temporal variations of the population of a
given city driven exclusively by a stochastic `natural part' (birth-death
 processes and other exogenous processes). The exogenous shocks $\eta$ are
 assumed to be a Gaussian noise, with average $m$ and variance $2\sigma^2$.

Starting from \eqref{eq-1city} as a minimal building block, we
consider a minimal model for the impact on population growth of
increasing accessibility and which consists of two cities, 1 and
2. Having in mind a large city connected to a small peripheral one, we
assume that city 2 is much smaller than city 1, $P_2 \ll P_1$ (City 1
that can in fact be considered as the whole world outside city 2). We
also assume that there is possibly a migration from city 1 to city 2
that is described by the rate $J$ (and we neglect the counterflow from
2 to 1). The evolution of populations 1 and 2 is then given in this
framework by
\begin{align}
\label{eq-2cities}
\frac{d P_1}{d t} &= \eta_1 P_1   \\
\frac{d P_2}{d t} &= \eta_2 P_2 + J P_1 \nonumber
\end{align}

In the case in which the cities are disconnected
($J = 0 $), they grow with the natural rate
\begin{align}
P_1(t) &= P_1(0)e^{\int^t \eta_1 (t')} \\
P_2(t) &= P_2(0)e^{\int^t \eta_2 (t')}, \nonumber
\end{align}
which on average implies (assuming that
both $\eta_1$ and $\eta_2$ have the same average $m$ and variance
$2 \sigma^2$)
\begin{align}
\langle P_1(t) \rangle = P_1(0)e^{(m + \sigma^2) t} \\
\langle P_2(t) \rangle = P_2(0)e^{(m + \sigma^2) t} , \nonumber
\end{align}
.

This will imply that both cities have the same average growth rate
\begin{equation}
g_1 = g_2 = \frac{d\langle P_{1(2)}\rangle}{\langle P_{1(2)} \rangle d t} = m + \sigma^2.
\end{equation}

The formal solutions to equations \eqref{eq-2cities} would be
\begin{align}
P_1(t) &= P_1(0) \left( e^{\int_0^t \eta_1 (t')}  \right) \\
P_2(t) &= P_2(0)e^{\int_0^t \eta_2 (t')} + \int_0^t J(t') P_1(t') e^{\int_{t'}^{t} \eta_2 (t'')}. \nonumber
\end{align}
where we consider the general case in which $J$ can depend on time,
and we use the fact that the equations for $P_1$ is decoupled.

We plan here to look at solutions of these equations
\eqref{eq-2cities}, for situations where $J$ is not constant in
time. The simplest scenario we have in mind a situation in which city
2 is at the beginning $t = t_0$ it gets coupled to the first city, and
then after some time the coupling goes again to zero.  Our scenario
would correspond to the following:
\begin{align}
J(t) &= 0 \ \ \text{for}\ \  t < t_0  \\
J(t) &= J \ \ \text{for}\ \  t_0 < t < t_0 + \delta t \nonumber \\
J(t) &= 0 \ \ \text{for}\ \  t > t_0 + \delta t \nonumber 
\end{align}
It is simple to compute the dynamics of the average values of
populations $\langle P_1 \rangle (t) $ and $\langle P_2 \rangle (t)$
that are simply
\begin{align}
\langle P_1 \rangle (t) &= P_1(0) e^{(m + \sigma^2) t} \\
\langle P_2 \rangle (t) &= P_2(0) e^{(m + \sigma^2) t} + \int_0^t J(t') \langle P_1(t') \rangle  e^{(m + \sigma^2) (t-t')}  \\
 &= P_2(0) e^{(m + \sigma^2) t} + P_1 (0)  e^{(m + \sigma^2) t }\int_0^t J(t')dt'.
\end{align}

In this case, it is sufficient to compute the integral $\int_0^t J(t')$ for our cases.
For case constant $J(t) = J$, it gives a contribution $J t$, for $J$ dependent on time we have
\begin{align}
\int_0^t J(t') &= 0 \ \ \text{for}\ \  t < t_0  \\
\int_0^t J(t') &= J (t - t_0)\ \ \text{for}\ \  t_0 < t < t_0 + \delta t \nonumber \\
\int_0^t J(t') &= J \delta t \ \ \text{for}\ \  t > t_0 + \delta t \nonumber . 
\end{align}

It is straightforward to show that the growth rate $g_2$ for the city 2 is
\begin{align}
g_2 &= m + \sigma^2 \ \ \text{for}\ \  t < t_0  \\
g_2 &= m + \sigma^2 \ \ \text{for}\ \  t > t_0 + \delta t \nonumber .
\end{align}

In the case $  t_0 < t < t_0 + \delta t$, we have
\begin{align}
  \frac{d\langle
  P_2(t)\rangle}{dt}=(m+\sigma^2)e^{(m+\sigma^2)t}\times\\
  \left[P_2(0)+P_1(0)J(t-t_0)+P_1(0)J/(m+\sigma^2)\right]
\end{align}
which leads to
\begin{align}
g_2=\frac{1}{\langle P_2(t)\rangle}  \frac{d\langle
  P_2(t)\rangle}{dt}&=m+\sigma^2+\frac{P_1(0)J}{P_2(0)+P_1(0)J(t-t_0)}\\
                 &=m+\sigma^2+\frac{\langle P_1(t)\rangle}{\langle P_2(t)}J
\end{align}
In the limit where the number of newcomers in city 2 is much less than the
population $P_2$: $P_1(0)J\delta t\ll P_2(0)$ we obtain
\begin{align}
\frac{\langle P_1(t)\rangle}{\langle P_2(t)\rangle}\approx
  \frac{P_1(0)}{P_2(0)}
\end{align}
which implies that the correction to the growth rate due to migration
is proportional to $J$.

In summary, we obtain for $g_2$ the following behavior
\begin{align}
g_2 &= m + \sigma^2 \ \ \text{for}\ \  t < t_0  \\
g_2 &= m + \sigma^2 + J \ \frac{\langle P_1(t) \rangle}{\langle P_2(t) \rangle}  \ \ \text{for}\ \  t_0 < t < t_0 + \delta t \nonumber \\
g_2 &= m + \sigma^2 \ \ \text{for}\ \  t > t_0 + \delta t \nonumber .
\end{align}
where we assumed that $J\langle P_1\rangle\delta t\ll P_2(0)$, or in
other words that the number of individuals moving from city 1 to city
2 is small compared to the population of city 2.

This model thus predicts an increase of growth rates that is linear in
$J$, and is inversely proportional to the population of the city.

\subsection{Testing the model}

The very simple model introduced in the previous sections provides us
a basic mechanism for which we can expect growth rate variations of cities in
the urban areas to depend linearly on the variation of accessibility
\begin{equation}
\Delta g = \kappa \Delta A .
\end{equation}

As we have seen however, a significant effect seems to be quantifiable
only for the few $0.5$ or $1\%$ of cities who experienced the largest
variation of accessibility. We decide here to restrict our attention
to this small set of cities which experienced some significant change
in accessibility, and test if such a linear dependence can indeed be
observable.

\begin{figure*}
  \begin{center}
        \includegraphics[scale=0.5]{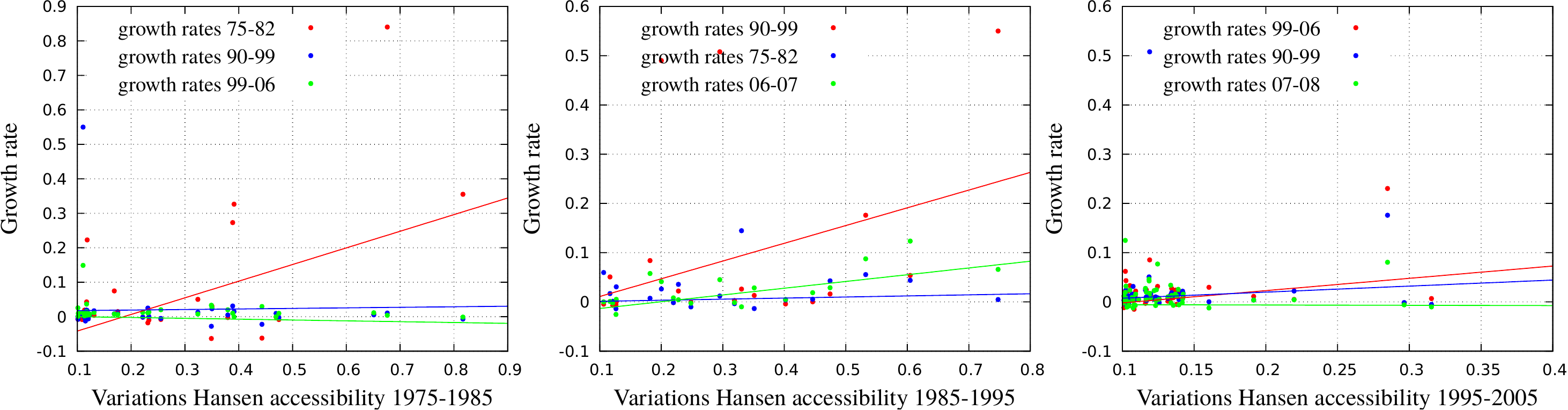}  
\caption{\textbf{Variations of Hansen accessibility and growth
    rates}. Growth rates distribution, when plotted against variations
  of accessibility show a weak but significative dependence, but only
  when the cities which experienced the largest accessibility
  variations are considered. }
\label{fig:A2-linear-fit}
\end{center}
\end{figure*}

In figure \ref{fig:A2-linear-fit} we show the scatter plot of all
cities in the top $1\%$ of accessibility variations (for the periods
1975-1985, 1985-1995 and 1995-2005). If we plot the growth rates for
these cities in different period, we see that a weak dependence, for
the growth rates in the relevant periods can be observed, while growth
rates in non relevant periods do not show some significant dependence.

%
%


\end{document}